\documentclass[conference,a4paper]{APSIPA2025}
\usepackage{amsmath}
\usepackage{amssymb}
\usepackage{graphicx}
\usepackage{booktabs}
\usepackage{threeparttable}
\usepackage{tabularx}
\usepackage[linesnumbered,ruled,vlined]{algorithm2e}

\usepackage{geometry}
\usepackage{fancyhdr}

\fancypagestyle{firststyle}{
  \fancyhf{}
  \fancyhead[C]{2025 Asia Pacific Signal and Information Processing Association Annual Summit and Conference (APSIPA ASC)}
}

\usepackage{multirow}
\usepackage{comment}
\usepackage{array}
\begin{document}

\title{Continual Audio Deepfake Detection via Universal Adversarial Perturbation}

\author{
\authorblockN{
Wangjie Li\authorrefmark{1},
Lin Li\authorrefmark{1} and Qingyang Hong\authorrefmark{2}
}

\authorblockA{
\authorrefmark{1}
School of Electronic Science and Engineering, Xiamen University, China}

\authorblockA{
\authorrefmark{2}
School of Informatics, Xiamen University, China\\
E-mail: liwangjie@stu.xmu.edu.cn, \{lilin, qyhong\}@xmu.edu.cn}
}
\maketitle
\thispagestyle{firststyle}
\pagestyle{fancy}

\begin{abstract}
  The rapid advancement of speech synthesis and voice conversion technologies has raised significant security concerns in multimedia forensics. Although current detection models demonstrate impressive performance, they struggle to maintain effectiveness against constantly evolving deepfake attacks. Additionally, continually fine-tuning these models using historical training data incurs substantial computational and storage costs. To address these limitations, we propose a novel framework that incorporates Universal Adversarial Perturbation (UAP) into audio deepfake detection, enabling models to retain knowledge of historical spoofing distribution without direct access to past data. Our method integrates UAP seamlessly with pre-trained self-supervised audio models during fine-tuning. Extensive experiments validate the effectiveness of our approach, showcasing its potential as an efficient solution for continual learning in audio deepfake detection.
\end{abstract}

\section{Introduction}
  In recent years, significant advancements in speech synthesis (TTS) and voice conversion (VC) technologies have made it increasingly difficult to distinguish between genuine and artificially generated speech~\cite{masood2023deepfakes,pmlr-v202-liu23f,Freevc}. These synthesized voices are often highly realistic, capable of deceiving human listeners with ease. While such technologies offer numerous beneficial applications, they also introduce severe security risks, including privacy violations, identity fraud and other malicious activities. As a result, there is a pressing need to advance audio deepfake detection to keep pace with these evolving threats.
  Community-driven initiatives, such as the ASVspoof Challenges~\cite{wu2015asvspoof,kinnunen2017asvspoof,todisco2019asvspoof,yamagishi2021asvspoof,wang2025asvspoof} and the Audio Deepfake Detection Challenges~\cite{yi2022add,yi2023add}, have played a pivotal role in driving progress in this field. Various techniques, including data augmentation\cite{tak2022rawboost,wang2024generalizable,zhang2024cpaug}, and multi-feature fusion\cite{wang23x_interspeech,guragain2024speech,pan24c_interspeech}, have been explored to enhance model generalization by extracting robust audio representations. Furthermore, fine-tuning pre-trained self-supervised learning models has significantly improved detection performance, achieving remarkable results on publicly available datasets~\cite{tak2022automatic,wang22_odyssey,wu2024spoofing}.

  As speech generation techniques continue to advance, detection models must evolve accordingly to counter emerging spoofing methods. Simply updating the model with newly collected spoofing data risks catastrophic forgetting, where previously learned patterns are partially lost. A straightforward approach to mitigate this issue is to retrain the model using both newly acquired and historical data. Nevertheless, this strategy is not only computationally and storage-intensive but also raises security concerns, such as data leakage.
  A more effective solution lies in continual learning, which enables the model to adapt to novel attack techniques while retaining knowledge from previous datasets. In the field of audio deepfake detection, few strategies about continual learning have been explored~\cite{ma21b_interspeech,zhang2023you,zhang2024remember}, and most existing approaches employ trainable gradient correction mechanisms to optimize model weights. Although these methods demonstrate effectiveness in mitigating catastrophic forgetting, applying gradient modifications to all neurons imposes constraints on the learning plasticity, potentially limiting its ability to adapt to new spoofing attacks.

  In this paper, we propose a novel framework that leverages Universal Adversarial Perturbation (UAP) to preserve historical knowledge in audio deepfake detection.
  UAP is a distinct type of imperceptible perturbation, specifically crafted to mislead deep learning models with high success rates~\cite{moosavi2017universal}. Originally introduced for image-related tasks~\cite{jetley2018friends,zhang2020understanding,sun2024continual}, UAP can be regarded as a feature that captures the primary data-space direction across class boundaries. Given that the distinctions between bona fide and spoofed audio are often subtle, UAP has the potential to serve as a key discriminative feature of spoofed audio relative to real audio. Our method leverages UAP and relevant data to approximate prior feature distributions and integrates them during fine-tuning stage. Instead of storing redundant spoofed data, our method requires retaining only a single UAP generated from the historical model. The combination of UAP and bona fide samples acts as pseudo-spoofed samples, effectively maintaining the distribution of the spoofed class without direct access to prior datasets. Our main contributions are as follows:
  \begin{itemize}
    \item We propose an effective training framework that integrates UAP into fine-tuning the pre-trained self-supervised audio model to preserve historical knowledge. To the best of our knowledge, we are the first to explore the applicability of UAP in audio deepfake detection. 
    \item We investigate the optimal strategy for incorporating UAP in continual learning and demonstrate that feature-level UAP outperforms waveform-level UAP with better retention of prior knowledge. 
    \item Experiments and visualizations validate the effectiveness of our approach, highlighting its potential as a robust and efficient solution for continual audio deepfake detection. 
    \end{itemize}
\begin{figure*}[t]
  \centering
  \includegraphics[width=\textwidth]{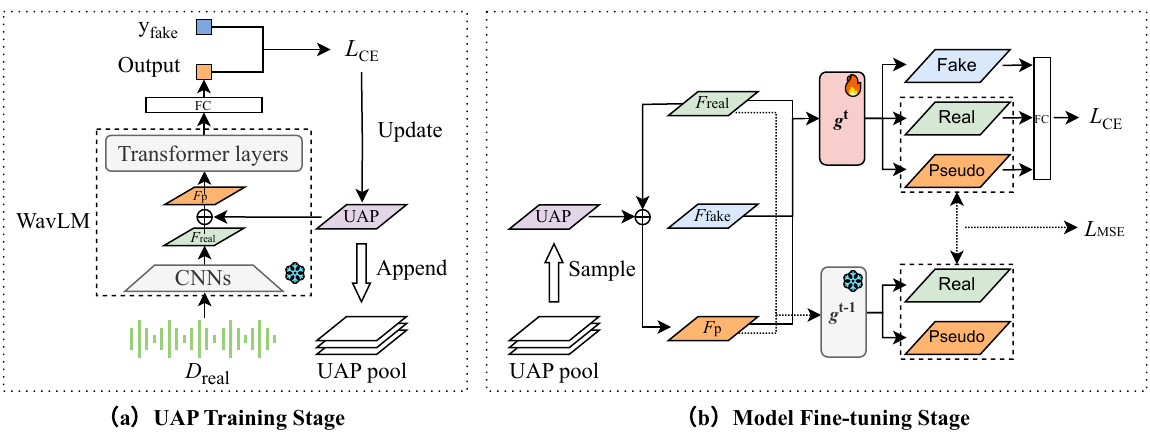}
  \caption{Overview of proposed training framework for continual audio deepfake detection. Subfigure (a) details the UAP training stage and subfigure (b) elucidates the model fine-tuning stage.}
  \label{fig:UAP}
\end{figure*}

\section{Method}
\subsection{Base training mechanism}
Continual audio deepfake detection aims to create a unified detector capable of handling a sequence of audio data from diverse attack types and domains. Formally, we denote training data available at the $t$-th stage as $D^t = \{D_{\text{real}}^t, D_{\text{fake}}^t\}$, where $D_{\text{real}}^t$ and $D_{\text{fake}}^t$ represent the bona fide and spoofed audio respectively. The audio deepfake detection system for the $t$-th stage is trained exclusively from dataset $D^t$. Data from the previous stages is no longer accessible. We introduce pre-trained self-supervised audio model to extract general acoustic features \(F \subset \mathbb{R^{B \times T \times D}}\) by the frozen convolutional neural network (CNN) layers. The extracted features consist of a batch of frame-level embeddings over time and frequency. The detection model $g$, consisting of Transformer layers followed by a fully connected layer, is optimized as a binary classifier using cross-entropy loss on dataset $D$, which is expressed as:
\begin{equation}
    L_{CE} = -\frac{1}{|D|}\sum_{x \in F, y \in D} y\log(g(x)) + (1-y)\log(1-g(x))
    \label{equation:eq0} 
\end{equation}
where $x \in \mathbb{R^{T \times D}}$ and $y \in \{0,1\}$ denote the extracted audio feature and corresponding label.

\subsection{UAP training stage}
As illustrated in Figure~\ref{fig:UAP} (a), we generate UAP using the model trained during the $t-1$ phase, and append it to a UAP pool.
For the classifier $g^{t-1}$ and bona fide audio feature \(x_{\text{real}}^{t-1} \in F_{\text{real}}^{t-1}\) extracted from bona fide subset \(D_{\text{real}}^{t-1}\) with label \(y_{\text{real}}=0\), our goal is to ascertain a perturbation vector \(p^{t-1} \in \mathbb{R^{T \times D}}\) capable of misleading the classifier into misclassifying real samples as fake. We target the vector \(p^{t-1}\) that conforms to:
\begin{equation}
  \begin{split}
  \text{Pred}(g^{t-1}&(x_{\text{real}}^{t-1}+p^{t-1})) = y_{\text{fake}}\\
  &\textbf{s.t.}\quad||p^{t-1}||_{\infty} \leq \epsilon
  \end{split}
\end{equation}
where \(\text{Pred}()\) converts the logit into a prediction value and adjusts the magnitude of the perturbation vector. The pseudo feature \(x^{t-1}_{p}\) is represented as:
\begin{equation}
  x^{t-1}_{p} = x_{\text{real}}^{t-1} + p^{t-1}
\end{equation}

We utilize the gradient descent technique to refine \(p^{t-1}\) by minimizing the entropy between the pseudo feature \(x^{t-1}_{p}\) and the fake label \(y_{\text{fake}}=1\), which is defined as:
\begin{equation}
  p^{t-1} = p^{t-1} - \alpha*\text{sgn}(\nabla_{p^{t-1}}\log(g^{t-1}(x^{t-1}_{p})))
  \label{equation:eq1}
\end{equation}
where \(\text{sgn}()\) represents the symbolic function and \(\alpha\) signifies the learning rate of \(p^{t-1}\). The perturbation is reiterated until \(x^{t-1}_{p}\) transgresses the decision boundary. The algorithm concludes when the number of successfully modified prediction outcomes surpasses the predetermined threshold \(\sigma\). After obtaining \(p^{t-1}\), we integrate it into the UAP pool in preparation for the next model fine-tuning phase.

\subsection{Model fine-tuning stage}
 Figure~\ref{fig:UAP} (b) demonstrates the integration of the UAP into the model fine-tuning stage. When fine-tuning the pre-tained audio model with newly collected dataset $D^t$ at the $t$-th phase, UAPs from previous stages up to $t-1$, denoted as $\{p^n\}_{n=1}^{t-1}$, are randomly sampled from the UAP pool at the start of each training iteration. Subsequently, pseudo features $F^n_{p}$ are created by adding $p^n$ to $F_{real}^t$ on an element-wise basis, formulated as $F^n_{p} = p^n + F_{real}^t$, and assigned the label $y_{fake}=1$. Then we combine \(F_{\text{p}}^{n}\) with \(F_{\text{real}}^{t}\) and \(F_{\text{fake}}^{t}\) to fine-tune the classifier. 

As parameters are dynamically optimized, the consistency of the pseudo-spoofed distribution may be compromised. To address this, we implement knowledge distillation between prior and current classifiers on the last Transformer layer. This maintains distillation based on the pseudo-spoofed feature $x^{t}_{p}$, which is formalized as:
\begin{equation}
  L^t_{p} =  \frac{1}{|F^n_{p}|}\sum_{x_{p}\in F^n_{p}}||g^t(x_{p}) - g^{t-1}(x_{p})||^2
  \label{equation:eq6}
\end{equation}

In real-world scenarios, bona fide audio tends to be more uniform than spoofed samples~\cite{One-ClassLearning,One-ClassKnowledge,kim24b_interspeech}. However, emerging spoofing methods can alter the distribution, impacting the generation of pseudo-spoofed samples. To mitigate this, we apply feature-based knowledge distillation, following the same paradigm as $L_p$. The previous classifier $g^{t-1}$ serves as the teacher, with a mean-squared loss function used as the regularization term. This can be articulated as:

\begin{equation}
  L^t_{r} = \sum_{x\in F_{real}^t}||g^t(x) - g^{t-1}(x)||^2
  \label{equation:eq7}
\end{equation}
\subsection{Loss function}
When new spoofing attacks arise at the $t$ stage, the overall loss function for fine-tuning model with UAP is defined as the sum of the base training loss and aforementioned distribution-preserving optimizations. This can be expressed as:
\begin{equation}
L^t = L^t_{CE} + \lambda(L^t_{MSE}) = L^t_{CE} + \lambda(L^t_{r}+L^t_{p})
\label{equation:final}
\end{equation}
where $\lambda$ is a constant determined based on the loss value of $L^t_{CE}$ to ensure they remain at the same order of magnitude.

\section{Experimental setup}
\subsection{Datasets and metrics}

Our experiments are conducted on three publicly available audio deepfake datasets, which are described as follows.

\textbf{ASVspoof 2019 LA}~\cite{todisco2019asvspoof} is one of the most commonly used English datasets in audio anti-spoof research. The training and development set share the same attack types including four TTS and two VC algorithms, while the evaluation set contains totally different attack types. The bona fide audio is collected from the VCTK corpus~\cite{veaux2017cstr}. 

\textbf{CFAD}~\cite{MA2024103122} is the first public Chinese standard dataset for audio deepfake detection under noisy and transcoding conditions. Twelve mainstream TTS techniques are used to generate spoofed audio. To simulate the real-world scenarios, three noise datasets and six codecs are considered for noise adding and audio transcoding.
\textbf{ASVspoof 5}~\cite{wang2025asvspoof} is the latest edition of the ASVspoof challenge series. Compared to previous challenges, the ASVspoof 5 database is built upon the MLS English dataset~\cite{pratap20_interspeech}, including crowdsourced data collected from a vastly greater number of speakers in diverse acoustic conditions. Attacks are generated and tested using surrogate detection models, while adversarial attacks and contemporary neural codecs are incorporated for the first time. 
To investigate the generalization ability of different methods, we also evaluate model performance on ASVspoof 2021 dataset~\cite{yamagishi2021asvspoof}. It is an influential evaluation dataset, sharing the same train and development subsets as ASVspoof 2019. Utterances of ASVspoof 2021 LA (21LA) dataset are transmitted over various channels while ASVspoof 2021 DeepFake (21DF) dataset collects numerous utterances processed with various lossy codecs used for media storage. Details are shown in Table~\ref{table:datasets}.
The Equal Error Rate (EER), which is widely used for audio deepfake detection, is applied to evaluate the experimental performance. 

\subsection{Model architecture}
We utilize the pre-trained self-supervised audio model WavLM~\cite{chenWavlm} as an efficient feature extractor for audio deepfake detection task. The architecture integrates CNN layers with Transformer encoders to extract speech features at multiple levels. We freeze the CNN layers to extract general acoustic features all the time. After extraction, these features are further processed by twelve Transformer encoders. Each Transformer block features sixteen attention heads and a hidden dimension of 768. Embedding from the last layer is then averaged and passed through the dropout with probability 0.2 and a fully connected layer to get the final prediction score.

\begin{table}
  \centering
  \caption{Summary of data statistics used in this study. \#Spks, \# Utts, \# Dur and \# Conds refer to the number of speakers, utterances, approximate duration and spoofing conditions, respectively. Division of train and test set is indicated by / .}
  \label{table:datasets}
  \resizebox{\linewidth}{!}{%
  \begin{tabular}{lcccc} 
  \toprule
  \multicolumn{1}{c}{Dataset} & \# Spks    & \# Utts          & \# Dur (h) & \# Conds  \\ 
  \hline
  ASVspoof 2019 LA            & 20 / 48   & 25,380 / 71,237   & 48 / 62      & 6 / 13   \\
  CFAD                        & 407 / 728 & 158,376 / 189,094 & 176 / 216       & 8 / 12   \\
  ASVspoof 5                  & 400 / 737 & 182,357 / 680,774 & 886 / 1345      & 8 / 16   \\ 
  \hline
  ASVspoof 2021 LA            & - / 48    & - / 181,566          & - / 130       & - / 13   \\
  ASVspoof 2021 DF            & - / 48    & - / 611,829          & - / 500       & - / 13   \\
  \bottomrule
  \end{tabular}
  }
  \end{table}

  \begin{table*}
    \centering
    \caption{The evaluation results of UAP with different sequence orders in terms of EER (\%). The \textbf{best} results are highlighted in bold, while the \underline{second-best} are underlined in each evaluation set.}
    \label{table:uap}
    \resizebox{\linewidth}{!}{%
    \begin{tabular}{cccccccccc} 
    \toprule
    \multirow{2}{*}{\textbf{ID}} & \multicolumn{3}{c}{\textbf{Initial Training Set}} & \multirow{2}{*}{\textbf{Sequence Order}} & \multirow{2}{*}{\textbf{UAP}} & \multicolumn{3}{c}{\textbf{Evaluation Set }} & \multirow{2}{*}{\textbf{Average}}  \\ 
    \cline{2-4}\cline{7-9}
                                 & ASVspoof2019LA & CFAD & ASVspoof 5                  &                                          &                               & ASVspoof2019LA & CFAD         & ASVspoof 5     &                                   \\ 
    \hline
    B1                           & \checkmark            &      &                             & $D_{1}$                                        &                               & 2.6          & 27.3         & 18.0           & 16.0                              \\
    B2                           &              & \checkmark    &                             & $D_{2}$                                        &                               & 14.7         & 5.0          & 23.1           & 14.3                              \\
    B3                           &              &      & \checkmark                           & $D_{3}$                                        &                               & 13.9         & 31.8         & 3.2            & 16.3                              \\ 
    \hline
    S1                           & \checkmark            &      &                             & $D_{1}$ $\rightarrow$ $D_{2}$                                  &                               & 8.5          & 4.8          & 21.9           & 11.7                              \\
    S2                           & \checkmark            &      &                             & $D_{1}$ $\rightarrow$ $D_{2}$                                  & \checkmark                             & 1.7          & \textbf{4.3} & 19.1           & \underline{8.3}                               \\
    S3                           & \checkmark            &      &                             & $D_{1}$ $\rightarrow$ $D_{2}$ $\rightarrow$ $D_{3}$                            &                               & 17.3         & 27.9         & \textbf{2.8}   & 16.0                              \\
    S4                           & \checkmark            &      &                             & $D_{1}$ $\rightarrow$ $D_{2}$ $\rightarrow$ $D_{3}$                             & \checkmark                             & 2.4          & 13.0         & \underline{9.5}            & \underline{8.3}                               \\
    S5                           &              &      & \checkmark                           & $D_{3}$ $\rightarrow$ $D_{2}$                                  &                               & 8.3          & \underline{4.5}          & 17.0           & 9.9                               \\
    S6                           &              &      & \checkmark                           & $D_{3}$ $\rightarrow$ $D_{2}$                                  & \checkmark                             & 6.2          & \underline{4.5}          & 13.0           & \textbf{7.9}                      \\
    S7                           &              &      & \checkmark                           & $D_{3}$ $\rightarrow$ $D_{2}$ $\rightarrow$ $D_{1}$                             &                               & \textbf{1.5} & 16.6         & 14.6           & 10.9                              \\
    S8                           &              &      & \checkmark                           & $D_{3}$ $\rightarrow$ $D_{2}$ $\rightarrow$ $D_{1}$                             & \checkmark                             & \underline{1.6}          & 13.1         & 10.7           & 8.5                               \\
    \bottomrule
    \end{tabular}
    }
    \end{table*}
\subsection{Implementation details} 
We applied normalization as pre-processing operation, which is calculated by subtracted mean and divided by standard deviation in each sample independently.
Fixed length of four seconds audio segments were used with zero-padding. 

When evaluating each audio, we make three four-second crops and score them independently. The final score for each sample is obtained by averaging individual scores.
To increase the data diversity, we involve noises and reverberation augmented data from the MUSAN and RIR corpus~\cite{snyder2015musan,ko2017study}, and adopt RawBoost~\cite{tak2022rawboost} of three noise algorithms. We applied each data augmentation on-the-fly with probability $0.5$ during the training process. 

We employed the pre-trained WavLM model from HuggingFace\footnote{huggingface.co/microsoft/wavlm-base}. Models were trained using Adam optimizer with hyperparameters $\beta = [0.9, 0.999]$ and weight decay $= 0$. While generating the UAP, we adjust the norm of perturbation ($\epsilon$) to $0.03$, set $\alpha$ as $0.0001$, and determine the successful threshold ($\sigma$) to be $0.8$. For fine-tuning process, we set the initial learning rate as $5e-5$ and hyperparameter $\lambda = 5$. We trained the model for $10$ epochs with an effective batch size of $128$ and selected the best-performing models on the development set. Training batches were formed using weighted random sampler with equal probabilities for classes. The complete framework is realized using PyTorch and runs on four NVIDIA A40 GPUs. All results are averaged over three runs.

\section{Results and analysis}
Since the sequence order is inherently agnostic for continual learning, we evaluate different training orders as shown in Table~\ref{table:uap}. \textit{Order1} follows a progression from simple to complex conditions, denoted as ASVspoof2019LA ($D_{1}$) $\rightarrow$ CFAD ($D_{2}$) $\rightarrow$ ASVspoof5 ($D_{3}$), while \textit{Order2} adopts the reverse order. Both orders involve cross-linguistic adaptation, encompassing a broad range of attack types and pronounced domain gaps.

\subsection{Baseline performance}
  Table~\ref{table:uap} shows that models trained on a specific dataset achieve high detection performance on their corresponding evaluation sets. However, their ability to generalize to out-of-domain data is significantly limited.  
  Specifically, as seen in baseline B1-B3, models trained on a particular domain (e.g., $D1$) show excellent performance on in-domain evaluation set but fail on evaluation sets from other domains ($D2$ and $D3$). These results validate the hypothesis that models optimized on one dataset struggle to maintain high generalization across diverse and unseen domains.
\subsection{Sequence fine-tuning with UAP}
  
  In subsequent experiments, we evaluated the impact of sequence fine-tuning (SFT) following different orders. Our findings indicate that while sequence fine-tuning improves performance on the target domain, it causes catastrophic forgetting, significantly reducing the model's performance on previous domains. In contrast, when UAP is incorporated during model fine-tuning, the model maintains strong performance on previous evaluation sets.
  Specifically, as shown in S3 and S4, sequence fine-tuning on target domain data results in a marked performance drop on prior domains. While leveraging UAP leads to an average improvement of 48\% relative to sequence fine-tuning. This demonstrates that UAP effectively preserves historical learned distribution and mitigates the loss of discriminative knowledge during optimization. 
  Similarly, S7 highlights that sequence fine-tuning, while adapting the model to the target domain, erases knowledge from prior datasets. On the other hand, the utility of UAP in experiment S8 successfully retains the historical spoofing distribution, achieving a 22\% relative enhancement in pooled evaluation performance. These findings from both sequence orders validate the effectiveness of our approach for continual audio deepfake detection.

  \begin{table}
    \centering
    \caption{EER (\%) comparison of systems with different UAP forms after sequence \textit{Order2} ($D_{3}$$\rightarrow$ $D_{2}$$\rightarrow$$D_{1}$).}
    \label{table:wavorFeat}
    \resizebox{\linewidth}{!}{%
    \begin{tabular}{cccccc} 
    \toprule
    Method        & 19LA         & 21LA         & 21DF         & CFAD          & ASVspoof5      \\ 
    \hline
    Joint Training & 2.9          & 3.9          & 8.6          & \textbf{4.9}           & \textbf{4.4}            \\ 
    SFT           & \textbf{1.5} & 4.6          & \textbf{7.8} & 16.6          & 14.6           \\
    UAP(waveform) & \underline{1.6}          & 4.3          & 8.2          & 14.4          & 13.8           \\
    UAP(feature)  & \underline{1.6}          & \textbf{3.8} & \underline{8.0}     & \underline{13.1} & \underline{10.7}  \\
    \bottomrule
    \end{tabular}
    }
  \end{table}
  \begin{figure}[t]
    \centering
    \includegraphics[width=\linewidth]{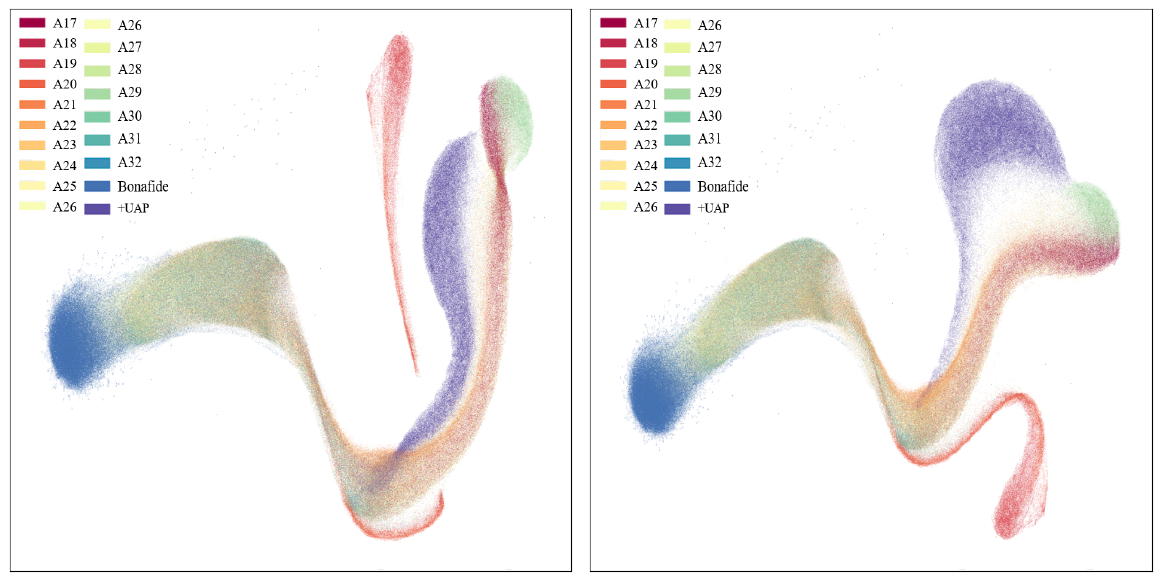}
    \caption{Visualization of embedding distribution on ASVspoof 5 evaluation set, plotted using UMAP~\cite{Umap} dimension reduction. Left: feature-level UAP.  Right: wavform-level UAP.}
    \label{fig:umap}
  \end{figure}
\subsection{Impact of different UAP forms }
  In our earlier experiments, we applied perturbations at the feature level to retain the model's previously acquired knowledge. Additionally, we investigated the impact of different forms of UAP during fine-tuning pre-trained models, specifically comparing feature-level and waveform-level UAP. Our objective was to determine which form of UAP more effectively preserves previously learned distribution while enhancing continual learning performance. Experiments in Table~\ref{table:wavorFeat} indicate that 
  although joint training on all datasets achieves strong average performance, it fails to attain optimal performance across all evaluation sets. Nevertheless, leveraging UAP consistently benefits the performance compared to sequence fine-tuning regardless of UAP forms. Moreover, feature-level UAP significantly outperforms waveform-level UAP in retaining prior knowledge of data distributions. This finding highlights the greater potential of feature-level UAP in ensuring better model stability and performance across different domains.

  To investigate the underlying reasons for performance discrepancy between two UAP forms, we visualize their effects on the ASVspoof 5 dataset in Figure~\ref{fig:umap}, where $A17$-$32$ represent eight attack types. The visualization shows that pseudo-spoofed samples created at feature level exhibit a closer distribution with spoofed samples than those at waveform level. We hypothesize that this is because the feature-level UAP captures less redundant details than waveform-level perturbation, making it more sensitive to the model decision. The effectiveness of our method relies on the inherent consistency of bona fide audio.

  Compared to raw waveforms, features extracted from a pre-trained model exhibit greater uniformity, thereby better preserving historical spoofing distribution and enhancing continual learning stability.

\section{Conclusions}
In this work, we introduce Universal Adversarial Perturbation (UAP) into audio deepfake detection to retain historical spoofing distribution while adapting to continuously emerging spoofing attacks. Specifically, we propose a training framework that effectively integrates UAP at the feature level when fine-tuning pre-trained self-supervised audio models. Additionally, knowledge distillation is adopted to preserve the distributional consistency of bona fide audio across different training stages. Extensive experiments and visualizations validate the effectiveness of our approach, highlighting its potential as a robust and efficient solution for continual learning in audio deepfake detection. In the future, we plan to explore more diverse adversarial perturbation strategies to better maintain model performance under complex and heterogeneous conditions in audio anti-spoofing.

\section*{Acknowledgment}
This work was supported in part by the National Natural Science Foundation
of China under Grants 62371407 and 62276220, and the Innovation of Policing
Science and Technology, Fujian province (Grant number: 2024Y0068).

\newpage
\bibliographystyle{IEEEtran}
\bibliography{MyBib.bib}

\begin{thebibliography}{10}
\providecommand{\url}[1]{#1}
\csname url@samestyle\endcsname
\providecommand{\newblock}{\relax}
\providecommand{\bibinfo}[2]{#2}
\providecommand{\BIBentrySTDinterwordspacing}{\spaceskip=0pt\relax}
\providecommand{\BIBentryALTinterwordstretchfactor}{4}
\providecommand{\BIBentryALTinterwordspacing}{\spaceskip=\fontdimen2\font plus
\BIBentryALTinterwordstretchfactor\fontdimen3\font minus \fontdimen4\font\relax}
\providecommand{\BIBforeignlanguage}[2]{{%
\expandafter\ifx\csname l@#1\endcsname\relax
\typeout{** WARNING: IEEEtran.bst: No hyphenation pattern has been}%
\typeout{** loaded for the language `#1'. Using the pattern for}%
\typeout{** the default language instead.}%
\else
\language=\csname l@#1\endcsname
\fi
#2}}
\providecommand{\BIBdecl}{\relax}
\BIBdecl

\bibitem{masood2023deepfakes}
M.~Masood, M.~Nawaz, K.~M. Malik, A.~Javed, A.~Irtaza, and H.~Malik, ``Deepfakes generation and detection: State-of-the-art, open challenges, countermeasures, and way forward,'' \emph{Applied intelligence}, vol.~53, no.~4, pp. 3974--4026, 2023.

\bibitem{pmlr-v202-liu23f}
H.~Liu, Z.~Chen, Y.~Yuan, X.~Mei, X.~Liu, D.~Mandic, W.~Wang, and M.~D. Plumbley, ``Audioldm: Text-to-audio generation with latent diffusion models,'' in \emph{International Conference on Machine Learning}.\hskip 1em plus 0.5em minus 0.4em\relax PMLR, 2023, pp. 21\,450--21\,474.

\bibitem{Freevc}
J.~Li, W.~Tu, and L.~Xiao, ``Freevc: Towards high-quality text-free one-shot voice conversion,'' in \emph{ICASSP 2023 - 2023 IEEE International Conference on Acoustics, Speech and Signal Processing (ICASSP)}, 2023, pp. 1--5.

\bibitem{wu2015asvspoof}
Z.~Wu, J.~Yamagishi, T.~Kinnunen, C.~Hanil{\c{c}}i, M.~Sahidullah, A.~Sizov, N.~Evans, M.~Todisco, and H.~Delgado, ``Asvspoof: the automatic speaker verification spoofing and countermeasures challenge,'' \emph{IEEE Journal of Selected Topics in Signal Processing}, vol.~11, no.~4, pp. 588--604, 2017.

\bibitem{kinnunen2017asvspoof}
T.~Kinnunen, M.~Sahidullah, H.~Delgado, M.~Todisco, N.~Evans, J.~Yamagishi, and K.~A. Lee, ``The asvspoof 2017 challenge: Assessing the limits of replay spoofing attack detection,'' in \emph{Interspeech 2017}, 2017, pp. 2--6.

\bibitem{todisco2019asvspoof}
M.~Todisco, X.~Wang, V.~Vestman, M.~Sahidullah, H.~Delgado, A.~Nautsch, J.~Yamagishi, N.~Evans, T.~H. Kinnunen, and K.~A. Lee, ``Asvspoof 2019: Future horizons in spoofed and fake audio detection,'' in \emph{Interspeech 2019}, 2019, pp. 1008--1012.

\bibitem{yamagishi2021asvspoof}
J.~Yamagishi, X.~Wang, M.~Todisco, M.~Sahidullah, J.~Patino, A.~Nautsch, X.~Liu, K.~A. Lee, T.~Kinnunen, N.~Evans \emph{et~al.}, ``Asvspoof 2021: accelerating progress in spoofed and deepfake speech detection,'' in \emph{ASVspoof 2021 Workshop-Automatic Speaker Verification and Spoofing Coutermeasures Challenge}, 2021.

\bibitem{wang2025asvspoof}
X.~Wang, H.~Delgado, H.~Tak, J.-w. Jung, H.-j. Shim, M.~Todisco, I.~Kukanov, X.~Liu, M.~Sahidullah, T.~Kinnunen \emph{et~al.}, ``Asvspoof 5: Design, collection and validation of resources for spoofing, deepfake, and adversarial attack detection using crowdsourced speech,'' \emph{Computer Speech \& Language}, p. 101825, 2025.

\bibitem{yi2022add}
J.~Yi, R.~Fu, J.~Tao, S.~Nie, H.~Ma, C.~Wang, T.~Wang, Z.~Tian, Y.~Bai, C.~Fan \emph{et~al.}, ``Add 2022: the first audio deep synthesis detection challenge,'' in \emph{ICASSP 2022 - 2022 IEEE International Conference on Acoustics, Speech and Signal Processing (ICASSP)}.\hskip 1em plus 0.5em minus 0.4em\relax IEEE, 2022, pp. 9216--9220.

\bibitem{yi2023add}
J.~Yi, J.~Tao, R.~Fu, X.~Yan, C.~Wang, T.~Wang, C.~Y. Zhang, X.~Zhang, Y.~Zhao, Y.~Ren \emph{et~al.}, ``Add 2023: the second audio deepfake detection challenge,'' \emph{arXiv preprint arXiv:2305.13774}, 2023.

\bibitem{tak2022rawboost}
H.~Tak, M.~Kamble, J.~Patino, M.~Todisco, and N.~Evans, ``Rawboost: A raw data boosting and augmentation method applied to automatic speaker verification anti-spoofing,'' in \emph{ICASSP 2022 - 2022 IEEE International Conference on Acoustics, Speech and Signal Processing (ICASSP)}.\hskip 1em plus 0.5em minus 0.4em\relax IEEE, 2022, pp. 6382--6386.

\bibitem{wang2024generalizable}
L.~Wang, L.~Yu, Y.~Zhang, and H.~Xie, ``Generalizable speech spoofing detection against silence trimming with data augmentation and multi-task meta-learning,'' \emph{IEEE/ACM Transactions on Audio, Speech, and Language Processing}, 2024.

\bibitem{zhang2024cpaug}
L.~Zhang, K.~A. Lee, L.~Zhang, L.~Wang, and B.~Niu, ``Cpaug: Refining copy-paste augmentation for speech anti-spoofing,'' in \emph{ICASSP 2024-2024 IEEE International Conference on Acoustics, Speech and Signal Processing (ICASSP)}.\hskip 1em plus 0.5em minus 0.4em\relax IEEE, 2024, pp. 10\,996--11\,000.

\bibitem{wang23x_interspeech}
C.~Wang, J.~Yi, J.~Tao, C.~Y. Zhang, S.~Zhang, and X.~Chen, ``Detection of cross-dataset fake audio based on prosodic and pronunciation features,'' in \emph{Interspeech 2023}, 2023, pp. 3844--3848.

\bibitem{guragain2024speech}
A.~Guragain, T.~Liu, Z.~Pan, H.~B. Sailor, and Q.~Wang, ``Speech foundation model ensembles for the controlled singing voice deepfake detection (ctrsvdd) challenge 2024,'' in \emph{2024 IEEE Spoken Language Technology Workshop (SLT)}.\hskip 1em plus 0.5em minus 0.4em\relax IEEE, 2024, pp. 774--781.

\bibitem{pan24c_interspeech}
Z.~Pan, T.~Liu, H.~B. Sailor, and Q.~Wang, ``Attentive merging of hidden embeddings from pre-trained speech model for anti-spoofing detection,'' in \emph{Interspeech 2024}, 2024, pp. 2090--2094.

\bibitem{tak2022automatic}
H.~Tak, M.~Todisco, X.~Wang, J.~weon Jung, J.~Yamagishi, and N.~Evans, ``Automatic speaker verification spoofing and deepfake detection using wav2vec 2.0 and data augmentation,'' in \emph{The Speaker and Language Recognition Workshop (Odyssey 2022)}, 2022, pp. 112--119.

\bibitem{wang22_odyssey}
X.~Wang and J.~Yamagishi, ``Investigating self-supervised front ends for speech spoofing countermeasures,'' in \emph{The Speaker and Language Recognition Workshop (Odyssey 2022)}, 2022.

\bibitem{wu2024spoofing}
H.~Wu, W.~Guo, Z.~Zhang, W.~Zhao, S.~Peng, J.~Zhang, and C.~M. Bank, ``Spoofing speech detection by modeling local spectro-temporal and long-term dependency,'' in \emph{Proc. Interspeech 2024}, 2024, pp. 507--511.

\bibitem{ma21b_interspeech}
H.~Ma, J.~Yi, J.~Tao, Y.~Bai, Z.~Tian, and C.~Wang, ``Continual learning for fake audio detection,'' in \emph{Interspeech 2021}, 2021, pp. 886--890.

\bibitem{zhang2023you}
X.~Zhang, J.~Yi, J.~Tao, C.~Wang, and C.~Y. Zhang, ``Do you remember? overcoming catastrophic forgetting for fake audio detection,'' in \emph{International Conference on Machine Learning}.\hskip 1em plus 0.5em minus 0.4em\relax PMLR, 2023, pp. 41\,819--41\,831.

\bibitem{zhang2024remember}
X.~Zhang, J.~Yi, C.~Wang, C.~Y. Zhang, S.~Zeng, and J.~Tao, ``What to remember: Self-adaptive continual learning for audio deepfake detection,'' in \emph{Proceedings of the AAAI Conference on Artificial Intelligence}, vol.~38, no.~17, 2024, pp. 19\,569--19\,577.

\bibitem{moosavi2017universal}
S.-M. Moosavi-Dezfooli, A.~Fawzi, O.~Fawzi, and P.~Frossard, ``Universal adversarial perturbations,'' in \emph{CVPR}, 2017, pp. 1765--1773.

\bibitem{jetley2018friends}
S.~Jetley, N.~Lord, and P.~Torr, ``With friends like these, who needs adversaries?'' \emph{NeurIPS}, vol.~31, 2018.

\bibitem{zhang2020understanding}
C.~Zhang, P.~Benz, T.~Imtiaz, and I.~S. Kweon, ``Understanding adversarial examples from the mutual influence of images and perturbations,'' in \emph{CVPR}, 2020, pp. 14\,521--14\,530.

\bibitem{sun2024continual}
K.~Sun, S.~Chen, T.~Yao, X.~Sun, S.~Ding, and R.~Ji, ``Continual face forgery detection via historical distribution preserving,'' \emph{International Journal of Computer Vision}, pp. 1--18, 2024.

\bibitem{One-ClassLearning}
Y.~Zhang, F.~Jiang, and Z.~Duan, ``One-class learning towards synthetic voice spoofing detection,'' \emph{IEEE Signal Processing Letters}, vol.~28, pp. 937--941, 2021.

\bibitem{One-ClassKnowledge}
J.~Lu, Y.~Zhang, W.~Wang, Z.~Shang, and P.~Zhang, ``One-class knowledge distillation for spoofing speech detection,'' in \emph{ICASSP 2024 - 2024 IEEE International Conference on Acoustics, Speech and Signal Processing (ICASSP)}, 2024, pp. 11\,251--11\,255.

\bibitem{kim24b_interspeech}
H.~M. Kim, K.~Jang, and H.~Kim, ``One-class learning with adaptive centroid shift for audio deepfake detection,'' in \emph{Interspeech 2024}, 2024, pp. 4853--4857.

\bibitem{veaux2017cstr}
C.~Veaux, J.~Yamagishi, K.~MacDonald \emph{et~al.}, ``Cstr vctk corpus: English multi-speaker corpus for cstr voice cloning toolkit,'' \emph{University of Edinburgh. The Centre for Speech Technology Research (CSTR)}, vol.~6, p.~15, 2017.

\bibitem{MA2024103122}
H.~Ma, J.~Yi, C.~Wang, X.~Yan, J.~Tao, T.~Wang, S.~Wang, and R.~Fu, ``Cfad: A chinese dataset for fake audio detection,'' \emph{Speech Communication}, vol. 164, p. 103122, 2024.

\bibitem{pratap20_interspeech}
V.~Pratap, Q.~Xu, A.~Sriram, G.~Synnaeve, and R.~Collobert, ``Mls: A large-scale multilingual dataset for speech research,'' in \emph{Interspeech 2020}, 2020, pp. 2757--2761.

\bibitem{chenWavlm}
S.~Chen, C.~Wang, Z.~Chen, Y.~Wu, S.~Liu, Z.~Chen, J.~Li, N.~Kanda, T.~Yoshioka, X.~Xiao, J.~Wu, L.~Zhou, S.~Ren, Y.~Qian, Y.~Qian, J.~Wu, M.~Zeng, X.~Yu, and F.~Wei, ``Wavlm: Large-scale self-supervised pre-training for full stack speech processing,'' \emph{IEEE Journal of Selected Topics in Signal Processing}, vol.~16, no.~6, pp. 1505--1518, 2022.

\bibitem{snyder2015musan}
D.~Snyder, G.~Chen, and D.~Povey, ``Musan: A music, speech, and noise corpus,'' \emph{arXiv preprint arXiv:1510.08484}, 2015.

\bibitem{ko2017study}
T.~Ko, V.~Peddinti, D.~Povey, M.~L. Seltzer, and S.~Khudanpur, ``A study on data augmentation of reverberant speech for robust speech recognition,'' in \emph{ICASSP 2017 - 2017 IEEE international conference on acoustics, speech and signal processing (ICASSP)}.\hskip 1em plus 0.5em minus 0.4em\relax IEEE, 2017, pp. 5220--5224.

\bibitem{Umap}
J.~Healy and L.~McInnes, ``Uniform manifold approximation and projection,'' \emph{Nature Reviews Methods Primers}, vol.~4, no.~1, p.~82, Nov 2024.

\end{thebibliography}

\end{document}